\documentclass[a4paper,leqno]{paper}

\newtheorem{definition}{Definition}

\newcommand{\ns}{\nobreakspace}
\usepackage{mathrsfs}
\usepackage[ansinew]{inputenc}
\usepackage{float}
\usepackage{afterpage}
\usepackage{amsmath}
\usepackage{amssymb}

\begin{document}

\begin{center}
\Large{\textbf{Some results on discrete eigenvalues for the
Stochastic Nonlinear Schr\"odinger Equation \\ in fiber optics}}
	\\[.5cm]
	\normalsize{Laura Prati and Luigi Barletti}	\\
   \vskip0.2cm

\vskip0.2cm
	\small{\textit{\textsf{Dipartimento di Matematica e Informatica ``U. Dini'' \\
Viale Morgagni 67/A, I-50134 Firenze, Italia}}}		

\end{center}

\begin{abstract}
We study a  stochastic Nonlinear Schr\"odinger Equation (NLSE), with additive white Gaussian noise, by means of the Nonlinear
Fourier Transform (NFT). 
In particular, we focus on the propagation of discrete eigenvalues along a focusing fiber. Since the stochastic 
NLSE is not exactly integrable by means of the NFT, then we use a perturbation approach, where 
we assume that the signal-to-noise ratio is high. 
The  zeroth-order perturbation leads to the deterministic NLSE while the  first-order perturbation allows to describe the statistics of the discrete eigenvalues.
This  is important to understand the properties of the channel for recently devised optical transmission techniques, where the information is encoded 
in the nonlinear Fourier spectrum.
\end{abstract}



\section{Introduction.}\label{sec.Introduction.}     %

The Nonlinear Schr\"odinger Equation (NLSE) governs the evolution of the complex envelope of signals propagating in optical fibers \cite{agr12,sec08}.  
The stochastic version of the NLSE also introduces the presence of optical noise, which is added to non-linear effects of propagation along the fiber. For a single-mode fiber, the stochastic NLSE in case of anomalous dispersion (focusing fiber) is
 \begin{equation} \label{eq.stochastic NLSE}
                  \frac{\partial q}{\partial z}(z,t)=\frac{i}{2} \frac{\partial^2 q}{\partial t^2}(z,t)+iq(z,t)|q(z,t)|^2+\eta(z,t)
 \end{equation} 
where $q(z,t)$ is the complex field envelope, $z\in \mathbb{R}^+$ is the normalized distance along the fiber, $t\in\mathbb{R}$ is time in a frame moving with the group velocity of the envelope, and $\eta(z,t)$ is the additive white Gaussian noise having zero mean \cite{pro83}:
\begin{equation}\label{eq.eta has zero mean}
                \langle\eta(z,t) \rangle =0\,.
\end{equation} 
Let  $\eta_1$ and $\eta_2$ denote the real and imaginary parts, respectively, of noise:
\begin{equation}\label{eq.real and imaginary part for eta}
                \eta(z,t)=\eta_1(z,t)+i\eta_2(z,t) \,.           
\end{equation}
We assume 
\begin{equation}\label{eq.correlation for eta} 
                \langle\eta_i(z,t) \eta_j(z',t') \rangle =D \delta(t-t')\delta(z-z') \delta_{ij}\,, \qquad i,j\in\{1,2\}
\end{equation}
where $D > 0$ is the spectral density of noise \cite{pap84}.
If the superscript $*$ represents complex conjugation, we have
\begin{equation}\label{eq.autocorrelation1 for eta and eta*} 
                \langle\eta(z,t) \eta^*(z',t') \rangle    =  \langle\eta^*(z,t) \eta(z',t') \rangle    =   2D \delta(t-t')\delta(z-z')
\end{equation}
and
\begin{equation}\label{eq.autocorrelation2 for eta and eta*}
                \langle\eta(z,t) \eta(z',t') \rangle      =  \langle\eta^*(z,t) \eta^*(z',t') \rangle  =   0  \,.
\end{equation}

The stochastic NLSE~\eqref{eq.stochastic NLSE} models the chromatic dispersion (first term in the right hand side), the Kerr nonlinearity (second term in the right hand side) and noise: these are physical aspects which seriously affect system performance during the propagation of the optical signal~\cite{agr12,sec08}.
 
 Equation~\eqref{eq.stochastic NLSE} has not been proven to be well-posed and is likely ill-posed~\cite{deb03}. Nevertheless, it is widely used in the physical and engineering literature~\cite{agr12,you14,civ15,nature16}. The problem can be mathematically fixed just with a cutoff in the spectrum of noise~\cite{bar15}. We do not introduce such a cutoff for simplicity of tractation.

The optical signal travels along the fiber moving from the transmitter to the receiver, which is possibly located tens of kilometers away.  It implies the risk of a strong distortion of the signal, making it difficult to decode the information at the receiver and limiting the transmission rate of the information itself.

Current optical networks mostly employ methodologies originally developed for linear channels, and nonlinearity has a detrimental effect on this kind of transmission. In particular, it has been estimated that the existing optical fiber technology is going to approach the so called \textit{nonlinear transmission limit}\ns\cite{mit01,ess08,ell10,ess10,ric10,kil11}. Linearization approaches to the NLSE can be found in \cite{sec08} and references therein and in \cite{bus15}. 
                                                     
Nowadays, part of the scientific community's efforts is devoted to develop technology for nonlinear and noisy transmissions. 

In this work, we approach the issue of nonlinearity by means of the Nonlinear Fourier Transform (NFT) \cite{has93,you14}, which is
a powerful tool to treat the (by definition) \textit{integrable} nonlinear partial differential equations, like the deterministic NLSE~\cite{akns73,akns74,libroAblowitz81,abl-clark91}:
\begin{equation} \label{eq.deterministic NLSE}
                  \frac{\partial q}{\partial z}(z,t)=\frac{i}{2} \frac{\partial^2 q}{\partial t^2}(z,t)+iq(z,t)|q(z,t)|^2\,.
 \end{equation}
The NFT maps the optical signal  (i.e.\ the solution of the NLSE) onto a set of scattering data -constituted by the continuous spectral function, the discrete eigenvalues and the discrete spectral function  associated to the deterministic NLSE\ns\eqref{eq.deterministic NLSE}- which evolve in a trivial manner along the fiber. 
On this ground, in \cite{has93} it was proposed to encode and transmit the information not on the optical signal itself, affected by dispersion and nonlinearity, but on the discrete eigenvalues associated to it, because they evolve linearly.   
This approach was named \textit{eigenvalue communication}. The choice of using the discrete spectrum for encoding the information has been explored by several authors, see \cite{oda04,mer12arxiv,ter13,you14-III,don15,har16}. Also the possibility of using the continuous spectrum has been approached, see \cite{pri13,pri14,le14,le15,nature16}, often under the name of \textit{nonlinear inverse synthesis}. 
 
We focus on the behavior of the discrete eigenvalues during the propagation along the fiber in the presence of additive white Gaussian noise in order to understand and describe their statistics and, as a direct consequence, the properties of the channel. Following 
\cite{you14-III,nature16}, we use a perturbation approach. Deterministic perturbation analysis for discrete eigenvalues can be found in  
\cite{kaup76,kar77,kaup78,kar81,kon94}, but a stochastic distribution of the spectral data is not faced. In \cite{kaz08} the special case of 
a noise-free channel  with a white Gaussian stochastic input is studied and the distributions of the spectral data are computed. 
In many works the statistics for the single-soliton case are studied, see \cite{fal04} and references therein. 
In \cite{you14-III} the perturbation of discrete eigenvalues is studied in order to simulate their statistics on a computer; the first-order 
variation of eigenvalues is obtained and analytical computations for the statistics are given explicitly for the one-soliton case.

In this work we examine in depth the non deterministic terms because the comprehension of their statistics lays the foundation for understanding the influence of noise on the received discrete spectrum. We assume that the signal-to-noise ratio is high ($\eta$ small) and we follow the perturbative approach of \cite{you14-III,nature16} to study analytically the statistics, in order to determine the hierarchy of perturbation equations for the propagation of eigenvalues. The zero-order perturbation leads to the deterministic NLSE~\eqref{eq.deterministic NLSE}.  

In Section~\ref{ssec.Basics on the Nonlinear Fourier Transform.} we briefly recall the NFT for the deterministic NLSE~\eqref{eq.deterministic NLSE} and its notation. In Section~\ref{ssec.Evolution of the scattering data.} we recall the evolution 
equations for the scattering data, also for small perturbations in the NFT. In Section\ns\ref{sec.Propagation of discrete eigenvalues.} we compute the output for discrete eigenvalues at the end of the fiber and the related statistics. As an example, in Section \ref{ssec.The one-soliton case.} we make explicit computations in the one-soliton case. In Section\ns\ref{sec.Conclusions.} we summarize some brief conclusions.
 
\section{Notation.}\label{sec.Notation.}                                             %
\subsection{Basics on the Nonlinear Fourier Transform.}                     %
               \label{ssec.Basics on the Nonlinear Fourier Transform.}       %

The Nonlinear Fourier Transform (NFT) is the forward step of the Inverse Scattering Transform (IST) technique, which was developed in the 1960s and later for solving certain nonlinear partial differential equations (PDEs), called \textit{integrable}  \cite{zab65,ggkm67,lax68,zs72,akns73,akns74},  
of the form
\begin{equation}\label{def.nonlinear PDE}
                q_z=K(q)
\end{equation}
where $q = q(z,t)$, $z \geq 0$, $t \in \mathbb{R}$, is sufficiently smooth  and vanishing as $t\to \pm \infty$ together with  all its derivatives, and $K$ is a differential operator acting on $q(z,\cdot)$.
 Numerical approaches to treat the NFT can be seen in \cite{wah13,wah15,civ15} where fast algorithms are described.

If $q$ is the solution of the deterministic NLSE \eqref{eq.deterministic NLSE}, which is integrable, one sets the so called \textit{Zakharov-Shabat (ZS) system}  for the evolution of the scattering data as:
\begin{equation}\label{eq.ZS system}
      \begin{cases}
                   &  v_{1t}(z,t)+i\, \zeta\, v_1(z,t)  = q(z,t) v_2(z,t) \\
                   &  v_{2t}(z,t)-i\, \zeta\, v_2(z,t)  = -q^*(z,t) v_1(z,t)
      \end{cases}
\end{equation} 
where $t\in\mathbb{R}$ is time, $z\in[0,\mathcal{L}]$ is the distance along a fiber of length $\mathcal{L}$,  $v(z,t)=(v_1(z,t),v_2(z,t))$ are the eigenfunctions, $\zeta\in\mathbb{C}$ are the eigenvalues, and $q^*(z,t)$ is the complex conjugate of $q(z,t)$, that  plays the role of a potential in the ZS system\ns\eqref{eq.ZS system}.
We assume that $q$  satisfies
\begin{equation}\label{eq.ipotesi q sommabile}
                \int_{-\infty}^{+\infty} |q(z,t)| dt < \infty
\end{equation}
and
\begin{equation}\label{eq.ipotesi q tende a zero}
                 q(z,t)\to 0  \qquad \textrm{for $|t|\to \infty$}\,.
\end{equation}
The eigenvalues $\zeta$ are generally complex:
\begin{equation}\label{eq.complex eigenvalue}
                        \zeta=\xi+i \mu \in \mathbb{C}       \qquad \textrm{with $\xi,\mu\in\mathbb{R}$;}
\end{equation}
when an eigenvalue is real, it is represented  by just $\xi\in\mathbb{R}$:
\begin{equation}\label{eq.real eigenvalue}
                       \zeta=\xi \in \mathbb{R} \,.
\end{equation}
Let us consider specific eigenfunctions  $\phi, \overline{\phi}, \psi, \overline{\psi}$ of the ZS system\ns\eqref{eq.ZS system} called \textit{Jost functions} (overbar does \textit{not} mean complex conjugation)  with the following asymptotic behavior for $|t|\to \infty$:
\begin{equation}
\begin{aligned}
 \phi(z,t;\xi)&=
      \begin{pmatrix}
           \phi_1(z,t;\xi)\\
           \phi_2(z,t;\xi)
       \end{pmatrix} 
 \to
     \begin{pmatrix}
           \textrm{e}^{-i\xi t}\\
           0
     \end{pmatrix}
\quad
&\textrm{as $t\to-\infty$} \\ 
\overline{\phi}(z,t;\xi)&=
      \begin{pmatrix}
           \overline{\phi}_1(z,t;\xi)\\
           \overline{\phi}_2(z,t;\xi)
       \end{pmatrix} 
 \to
    \begin{pmatrix}
           0\\
           -\textrm{e}^{i\xi t}
     \end{pmatrix}
\quad
&\textrm{as $t\to-\infty$}\\
\psi(z,t;\xi)&=
      \begin{pmatrix}
           \psi_1(z,t;\xi)\\
           \psi_2(z,t;\xi)
       \end{pmatrix} 
 \to
     \begin{pmatrix}
           0\\
           \textrm{e}^{i\xi t}
     \end{pmatrix}
\quad
&\textrm{as $t\to+\infty$}\\
 \overline{\psi}(z,t;\xi)&=
      \begin{pmatrix}
           \overline{\psi}_1(z,t;\xi)\\
           \overline{\psi}_2(z,t;\xi)
       \end{pmatrix} 
 \to
     \begin{pmatrix}
           \textrm{e}^{-i\xi t}\\
           0
     \end{pmatrix}
\quad
&\textrm{as $t\to+\infty$}
\end{aligned}
\end{equation}
The right and left solutions are related by the \textit{Jost coefficients} $a,b,\overline{a},\overline{b}$
\begin{equation}
                       \begin{aligned}
                                           &\psi(z,t;\xi)                = - a(\xi,z) \overline{\phi}(z,t;\xi)
                                                                                + \overline{b}(\xi,z) {\phi}(z,t;\xi)  \\
                                           &\overline{\psi}(z,t;\xi) = \overline{a}(\xi,z) {\phi}(z,t;\xi)
                                                                                + {b}(\xi,z) \overline{\phi}(z,t;\xi)
    \end{aligned}
\end{equation}  
where 
\begin{equation} 
                      a(\xi,z) \overline{a}(\xi,z) + {b}(\xi,z) \overline{b}(\xi,z) = 1.
\end{equation}

Functions $a,\overline{a}$ can be analytically extended (in $\xi$) into the upper and lower half complex plane respectively \cite{akns74}. 
Therefore, in addition to the \textit{continuous spectrum} ($\xi\in\mathbb{R}$), there is also the \textit{discrete spectrum}, 
that occurs for $\zeta_k\in\mathbb{C}^+$ such that 
\begin{equation}\label{eq.zeros of a}
a(\zeta_k,z)=0 \qquad k=1,2,\dots,N
\end{equation}
($N$ is finite, see \cite{akns74}). 
If $\zeta_k\in\mathbb{C}^+$ is an eigenvalue, also $\zeta_k^*\in\mathbb{C}^-$ is a discrete eigenvalue in the lower half plane, therefore one omits to consider the zeros of $\overline{a}$ and  limits the study of discrete eigenvalues to the upper half plane.
The Jost functions $\phi,\psi$ become linearly dependent \cite{akns74}:
\begin{equation}\label{eq.Jost functions discrete}
 \phi(z,t;\zeta_k)=b(\zeta_k,z)  \psi(z,t;\zeta_k) \qquad \textrm{for $k=1,2,\dots,N$}.
\end{equation}

The asymptotic behavior of the Jost functions for discrete eigenvalues $\zeta_k\in\mathbb{C}^+$ is given by  
\begin{equation}\label{eq.asymptotic behavior discrete phi -infty }
\begin{aligned}\phi(z,t;\zeta_k)&=
      \begin{pmatrix}
           \phi_1(z,t;\zeta_k)\\
           \phi_2(z,t;\zeta_k)
       \end{pmatrix} 
 \to
     \begin{pmatrix}
           \textrm{e}^{-i\zeta_k t}\\
           0
     \end{pmatrix}
\quad
&\textrm{as $t\to-\infty$}\\
 \psi(z,t;\zeta_k)&=
      \begin{pmatrix}
           \psi_1(z,t;\zeta_k)\\
           \psi_2(z,t;\zeta_k)
       \end{pmatrix} 
 \to
     \begin{pmatrix}
           0\\
           \textrm{e}^{i\zeta_k t}
     \end{pmatrix}
\quad
&\textrm{as $t\to+\infty$}
\end{aligned}
\end{equation}

\begin{definition}\label{def.scattering data}
                       The set of \textit{scattering data} is given by        
                   \begin{equation}\label{eq.def scattering data}
                                S_+:=\{\rho(\xi),\xi \in\mathbb{R}\}\cup \{\zeta_k\in\mathbb{C}^+,C_k\}_{k=1}^N  
                   \end{equation}
                       where
                   \begin{equation}\label{eq.def continuous spectral function}
                               \rho(\xi,z):=\frac{b(\xi,z)}{a(\xi,z)},  \qquad \textrm{for $\xi\in\mathbb{R}$},
                  \end{equation}   
                       is called the \textit{continuous spectral function}, defined on the continuous spectrum   
                       $\xi\in\mathbb{R}$, $\zeta_k$  are the \textit{discrete eigenvalues} in the upper half plane  
                       $\mathbb{C}^+$, and 
                 \begin{equation}\label{eq.def discrete spectral function}
                                C_k:=\frac{b(\zeta_k,z)}{a_\zeta(\zeta_k,z)}, \qquad \textrm{for $k=1,2,\dots,N$},
                 \end{equation}
                       is called the \textit{discrete spectral function}.
 \end{definition}
 The NFT maps the nonlinear PDE one aims at resolving into its set of scattering data. 

\subsection{Evolution of the scattering data and perturbations.}\label{ssec.Evolution of the scattering data.} %

In order to give the evolution equations for the scattering data, we are going to use the compact notation of \cite{kaup76}.
\begin{definition}\label{def.Kaup operator}
If $q(z,t)$ satisfies an evolution equation
\begin{equation}
   q_z=K(q)
\end{equation} 
where $K$ nonlinear operator acting on $q$ as a function of $t$, the Kaup operator $I[u,v;\zeta]$ acts on $u\ns=\ns(u_1,u_2),v=(v_1,v_2)$ like
\begin{equation}\label{eq.def_I}
                        I[u,v;\zeta]:=\!\!\int_{-\infty}^{+\infty}  \!\!\!
                                       \left[ iq_z(z,t) u_2(z,t;\zeta)v_2(z,t;\zeta)
                                                          +i q^*_z(z,t) u_1(z,t;\zeta)v_1(z,t;\zeta)
                                                    \right] dt.
\end{equation}
\end{definition}
Notice that the operator $I$ is bilinear. With this notation, the evolution of scattering coefficients $a,b$ and scattering data $\rho,\zeta_k
$ is given by \cite{kaup76}
\begin{align}
 & i a_z(\zeta,z)= I[\phi,\psi;\zeta], \\
 & i b_z(\zeta,z)= -I[\phi,\overline{\psi};\zeta], \\
 & i \rho_z(\xi,z)=-\frac{1}{a^2(\xi,z)} I[\phi,\phi;\xi], \\  \label{AAA}
 & i \frac{d}{dz}\zeta_{k}(z)=-C_k I[\psi,\psi;\zeta_k],    \qquad \textrm{for $k=1,2,\dots,N$,}
\end{align}
where $\phi$, $\psi$ and $\overline{\psi}$ are Jost functions of the ZS system associated to $K$ 
(we omit the complicated evolution of $C_k$ because not useful here).  When $q$ satisfies the 
deterministic NLSE \eqref{eq.deterministic NLSE}, one uses index $0$ for the Kaup operator $I_0[u,v;\zeta]$
and the evolution equations simplify to:
\begin{align}
       & i a_z(\zeta,z)= 0, \label{eq.evolution of a exactly solvable}\\
       & i b_z(\zeta,z)= -2\zeta^2b(\zeta,z), \\
       & i \rho_z(\xi,z)=-2\xi^2\rho(\xi,z), \\        
       & i \frac{d}{dz}\zeta_k(z)=0, \qquad \textrm{for $k=1,\dots,N$,} \\
       & i C_{k,z}(z)= -2 \zeta_k^2 C_k(z), \qquad \textrm{for $k=1,\dots,N$.} \label{eq.Ck exactly solvable}
\end{align}
The evolution is trivial and  the number of the discrete eigenvalues, which are zeros of $a$, is  preserved.

In \cite{kaup76,kaup78} a perturbation theory is developed for $q$ satisfying
\begin{equation}\label{eq.PDE q perturbed}
              q_z(z,t)=K_0(q)+\sigma K_1(q),
\end{equation}
with $K_0$ nonlinear operator for an exactly integrable case, like the deterministic NLSE \eqref{eq.deterministic NLSE}, $K_1$ operator for the perturbation and $\sigma$ a small parameter. Briefly, for the perturbation of the deterministic NLSE \eqref{eq.deterministic NLSE}, one has
\begin{equation}\label{eq.I_perturbed}
            I[u,v;\zeta]=I_0[u,v;\zeta]+\sigma I_1[u,v;\zeta],
\end{equation}
where
\begin{multline}
\label{I1def}
        I_1[u,v;\zeta]  :=
        \\
        \int_{-\infty}^{+\infty} \left[
                                            i K_1(q) u_2(z,t;\zeta)v_2(z,t;\zeta)
                                           +i [K_1(q)]^* u_1(z,t;\zeta)v_1(z,t;\zeta) \right] dt
\end{multline}
should be understood as computed for $u,v$  in the unperturbed problem (zero-order perturbation), therefore known  (full information).

This framework can therefore be used in order to treat the stochastic NLSE \eqref{eq.stochastic NLSE} for the propagation of signal along the fiber \cite{nature16}
by interpreting $K_1$ as the additive noise term.

\section{Propagation of discrete eigenvalues.}\label{sec.Propagation of discrete eigenvalues.}

 Following 
\cite{you14-III,nature16}, we consider a small noise  $\eta$, compared with the signal power, and apply the NFT perturbation theory to obtain a stochastic description of the channel. In \cite{nature16} the stochastic evolution for the continuous spectral function is studied. We are going to study the evolution for the discrete 
eigenvalues following Kaup's notation. A similar perturbation approach to obtain the evolution equation for discrete eigenvalues is given in 
\cite{you14-III}, where the perturbation of discrete eigenvalues is studied in order to simulate their statistics on a computer and analytical computations are given explicitly only for the one-soliton case. By means of the Kaup notation, we not only solve the evolution equation for discrete eigenvalues, but  we are also able (in principle) to give a complete description of the statistics of the channel for the multi-solitonic case, corresponding to an arbitrary number $N$ of eigenvalues in the upper half complex plane.

From \eqref{AAA} we have that the equation for the evolution of the discrete eigenvalues $\zeta_k=\zeta_k(z)$, for $k=1,2,\dots,N$ is therefore given by:
\begin{equation}\label{eq.evolution of zeta splitted}
 i \frac{d}{dz} \zeta_k(z)= -C_k(z)  I_1^\textrm{noise}[\psi,\psi;\zeta_k]\,
\end{equation}
where $ C_k(z)$ is defined in Equation\ns\eqref{eq.def discrete spectral function} 
and, according to \eqref{I1def},
\begin{multline}\label{eq.def I_1noise}
 I_1^\textrm{noise}[u,v;\zeta]:=
 \\
 \int_{-\infty}^{+\infty} \left[i \eta(z,t) u_2(z,t;\zeta)v_2(z,t;\zeta)
                                                                                              +i\eta^*(z,t)u_1(z,t;\zeta)v_1(z,t;\zeta)
                                                                                        \right] dt.
\end{multline}
 Equation\ns\eqref{eq.evolution of zeta splitted} can be written more explicitly as
\begin{equation}\label{eq.evolution discrete eigenvalues noisy case}
  \frac{d}{dz} \zeta_k(z)=-C_k(z)
 \! \int_{-\infty}^{+\infty} \left[ \eta(z,t) \psi^2_2(z,t;\zeta_k)
                                                                                              +\eta^*(z,t) \psi^2_1(z,t;\zeta_k)
                                                                                        \right] dt
\end{equation}
or, concisely,
\begin{equation}\label{eq.evolution of zetak with Delta}
       \frac{d}{dz} \zeta_k(z)=\Delta(\zeta_k,z)
\end{equation} 
where
\begin{equation}\label{eq.def of Delta}
       \Delta(\zeta_k,z):= -C_k(z)
  \int_{-\infty}^{+\infty} \left[ \eta(z,t) \psi^2_2(z,t;\zeta_k)
                                                                                        + \eta^*(z,t) \psi^2_1(z,t;\zeta_k)
                                                                \right] dt  .
\end{equation}
Note, moreover, that
\begin{equation}
 C_k(z) =C_k(\zeta_k(0),z) = \frac{b(\zeta=\zeta_k(0),z)}{a_\zeta(\zeta=\zeta_k(0),z)},
\end{equation}
since the Jost functions and the Jost coefficients should be understood as those of the 
unperturbed problem and are assumed to be known from the leading order. 

Let us compute the statistics of noise $\Delta(\zeta_k,z)$ to completely determine it.
The mean is null:
\begin{equation}\label{eq.mean value of Delta}
\langle\Delta(\zeta_k,z) \rangle 
                         =0
\end{equation}
because $ \langle\eta(z,t) \rangle =0$, see Equation\ns\eqref{eq.eta has zero mean}.
This implies that
\begin{equation}
 \frac{d}{dz}\langle\zeta_k(z) \rangle =0
\end{equation}
and then,  from Equations\ns\eqref{eq.evolution of zetak with Delta}  and \eqref{eq.mean value of Delta},
\begin{equation}
 \langle\zeta_k(z) \rangle =\langle\zeta_k(0) \rangle =\zeta_k(0).
\end{equation}
Hence, the mean value of each discrete eigenvalue is constant and equals its value at $z=0$.

The autocorrelations to be evaluated are $\langle\Delta(\zeta_k,z)\Delta^*(\zeta_j,z')\rangle$ and $\langle\Delta(\zeta_k,z)\Delta(\zeta_j,z')\rangle$ for $\zeta_k\neq\zeta_j,z\neq z'$ . For the first one, one obtains
\begin{equation*}
\begin{aligned}
\langle\Delta(\zeta_k,z)\Delta^*(\zeta_j,z') \rangle &=C_k(z)C_j^*(z') \times \\
 \times  \int_{-\infty}^{+\infty} \!\!\! dt
         \int_{-\infty}^{+\infty} \!\!\! dt'
              &  \left[ 2D \delta(t-t')\delta(z-z')  \psi^2_2(z,t;\zeta_k)\psi^{*2}_2(z',t';\zeta_j)\right. +\\
             &   \left.  +2D \delta(t-t')\delta(z-z') \psi^2_1(z,t;\zeta_k)\psi^{*2}_1(z',t';\zeta_j)
                \right],
\end{aligned}
\end{equation*}
where $D > 0$ is the spectral density of noise, see Equation\ns\eqref{eq.correlation for eta};  hence
\begin{equation*}
\begin{aligned}
\langle\Delta&(\zeta_k,z)\Delta^*(\zeta_j,z') \rangle =C_k(z)C_j^*(z')  2D \delta(z-z')\times\\
 &\times\int_{-\infty}^{+\infty}
                \left[   \psi^2_2(z,t;\zeta_k)\psi^{*2}_2(z',t;\zeta_j)
                 + \psi^2_1(z,t;\zeta_k)\psi^{*2}_1(z',t;\zeta_j)
                \right] dt,                     
\end{aligned}
\end{equation*}
and eventually, because of the presence of the term $\delta(z-z')$
\begin{equation}
\begin{aligned}
\langle\Delta&(\zeta_k,z)\Delta^*(\zeta_j,z') \rangle =C_k(z)C_j^*(z) 2D \delta(z-z')\times\\
 &\times\int_{-\infty}^{+\infty}
                \left[   \psi^2_2(z,t;\zeta_k)\psi^{*2}_2(z,t;\zeta_j)
                 + \psi^2_1(z,t;\zeta_k)\psi^{*2}_1(z,t;\zeta_j)
                \right] dt .                      
\end{aligned}
\end{equation}
Briefly, it is
\begin{equation}\label{eq.autocorrelation1 for Delta final}
       \langle\Delta(\zeta_k,z)\Delta^*(\zeta_j,z') \rangle = 2D\, C(z;\zeta_k,\zeta_j) \delta(z-z')
\end{equation}
where
\begin{multline}\label{eq.def of C}
       C(z;\zeta_k,\zeta_j):=C_k(z)C_j^*(z)  \times
       \\
     \times
       \int_{-\infty}^{+\infty} 
                  \left[   \psi^2_2(z,t;\zeta_k)\psi^{*2}_2(z,t;\zeta_j)
                           + \psi^2_1(z,t;\zeta_k)\psi^{*2}_1(z,t;\zeta_j)
                  \right] dt.
\end{multline}
Similarly
\begin{equation}\label{eq.autocorrelation2 for Delta final}
     \langle\Delta(\zeta_k,z)\Delta(\zeta_j,z') \rangle = 2D\, E(z;\zeta_k,\zeta_j) \delta(z-z')
\end{equation}
where
\begin{multline}\label{eq.def of E}
       E(z;\zeta_k,\zeta_j):=C_k(z)C_j(z)\times
       \\
 \times
       \int_{-\infty}^{+\infty}
                     \left[   \psi^2_2(z,t;\zeta_k)\psi^{2}_1(z,t;\zeta_j)
                              + \psi^2_1(z,t;\zeta_k)\psi^{2}_2(z,t;\zeta_j)
                     \right] dt
\end{multline}
The autocorrelation functions given in Equations\ns\eqref{eq.autocorrelation1 for Delta final} and\ns\eqref{eq.autocorrelation2 for Delta final} do make sense, because integrals in Equations\ns\eqref{eq.def of C} and\ns\eqref{eq.def of E} both converge because of the decay properties of the Jost functions \eqref{eq.asymptotic behavior discrete phi -infty }.  Indeed, the integrands are continuous functions and they decay exponentially where the  interval of integration goes to infinity. For example we show the convergence of  the integral defining $C(z;\zeta_k,\zeta_j)$ in Equation\ns\eqref{eq.def of C}. Consider an arbitrary $T>0$. Then we have: \begin{multline}  
C_k(z)C_j^*(z)
\int_{-\infty}^{-\frac{T}{2}} 
                \big[  \psi_2^2(z,t;\zeta_k)\psi_2^{*2}(z,t;\zeta_j) 
             +\psi_1^2(z,t;\zeta_k)\psi_1^{*2}(z,t;\zeta_j)
                \!\big]\, dt \\
=C_k(z)C_j^*(z) {b}^{-2}(\zeta_k,z) {b}^{*-2}(\zeta_j,z)
                \int_{-\infty}^{-\frac{T}{2}} \textrm{e}^{-2i(\zeta_k-\zeta_j^*) t}  dt+ O\left(\frac{1}{T}\right)
\\
=-\frac{C_k(z)C_j^*(z)}{{b}^{2}(\zeta_k,z) {b}^{*2}(\zeta_j,z)}\;
                     \frac{1}{2i(\zeta_k-\zeta_j^*)}\;
                     \textrm{e}^{i(\zeta_k-\zeta_j^*) T}+ O\left(\frac{1}{T}\right)
\end{multline}
where $\zeta_k=\alpha_k+i\beta_k$ with $\beta_k\in\mathbb{R}^+$ and  
$\zeta_j=\alpha_j+i\beta_j$ with $\beta_j\in\mathbb{R}^+$, then  
$i(\zeta_k-\zeta_j^*)$ 
has   
a negative real part. 
Similarly, on the right we have: 
\begin{multline}
C_k(z)C_j^*(z) 
\int_{\frac{T}{2}}^{+\infty}
                \big[  \psi_2^2(z,t;\zeta_k)\psi_2^{*2}(z,t;\zeta_j) + 
                +\psi_1^2(z,t;\zeta_k)\psi_1^{*2}(z,t;\zeta_j)
                \big]\, dt 
\\=  -C_k(z)C_j^*(z)\frac{1}{2i(\zeta_k-\zeta_j^*)}\;
                   \textrm{e}^{i(\zeta_k-\zeta_j^*) T}+ O\left(\frac{1}{T}\right) \,.  
\end{multline}

The convergence of $E(z;\zeta_k,\zeta_j)$ can be shown easily in a similar way.


\subsection{The channel  output.}\label{ssec.The channel  output.}  %

Let now $\zeta_k(\mathcal{L})$ be the channel output, defined as the solution of Eq.\eqref{eq.evolution of zetak with Delta}  
at the receiver located at distance $z=\mathcal{L}$.
 Equation \eqref{eq.evolution of zetak with Delta}  gives
\begin{equation}
 \zeta_k(z)=\zeta_k(0)+\int_0^z \Delta(\zeta_k(0),z')dz'
\end{equation}
therefore the channel output is
\begin{equation}
 \zeta_k(\mathcal{L})=\zeta_k(0)+\int_0^\mathcal{L} \Delta(\zeta_k(0),z)dz
\end{equation}
or shortly
\begin{equation}
 \zeta_k(\mathcal{L})=\zeta_k(0)+N(\zeta_k),
\end{equation}
where
\begin{equation}
 N(\zeta_k):=\int_0^\mathcal{L} \Delta(\zeta_k,z)dz\,.
\end{equation}

Let us compute the statistics of $N(\zeta_k)$. The mean is
\begin{equation}
 \langle N(\zeta_k)\rangle=0
\end{equation}
because of Equation\ns\eqref{eq.mean value of Delta}. Then we compute the two autocorrelations 
$ \langle N(\zeta_k)N^*(\zeta_j)\rangle$ and  $\langle N(\zeta_k)N(\zeta_j)\rangle$ 
for $\zeta_k\neq\zeta_j$. The first one is
\begin{equation}
 \begin{aligned}
    \langle N(\zeta_k)N^*(\zeta_j)\rangle
                      & = \int_0^\mathcal{L} dz \int_0^\mathcal{L} dz'\, 
                           \langle \Delta(\zeta_k,z) \Delta^*(\zeta_j,z')\rangle\\
                      & = \int_0^\mathcal{L} dz \int_0^\mathcal{L} dz'\, 
                          2D\, C(z;\zeta_k,\zeta_j) \delta(z-z')
\end{aligned}
\end{equation}
where we have used Equation\ns\eqref{eq.autocorrelation1 for Delta final},  and eventually
\begin{equation}\label{eq.autocorrelation1 for N discrete}
    \langle N(\zeta_k)N^*(\zeta_j)\rangle = 2D \int_0^\mathcal{L}   C(z;\zeta_k,\zeta_j)\,dz\,.
\end{equation}
Similarly, because of  Equation\ns\eqref{eq.autocorrelation2 for Delta final}, it is
\begin{equation}
 \begin{aligned}
    \langle N(\zeta_k)N(\zeta_j) \rangle
                      & = \int_0^\mathcal{L} dz \int_0^\mathcal{L} dz'\, 
                           \langle \Delta(\zeta_k,z) \Delta(\zeta_j,z')\rangle\\
                      & = \int_0^\mathcal{L} dz \int_0^\mathcal{L} dz'\, 
                          2D\, E(z;\zeta_k,\zeta_j) \delta(z-z')\\
\end{aligned}
\end{equation}
or
\begin{equation}\label{eq.autocorrelation2 for N discrete}
    \langle N(\zeta_k)N(\zeta_j)\rangle = 2D \int_0^\mathcal{L}  E(z;\zeta_k,\zeta_j)\,dz\,.
\end{equation}
Both the autocorrelations can be therefore computed with the knowledge of Jost coefficients and Jost functions 
in the unperturbed case.

\subsection{The one-soliton case.}\label{ssec.The one-soliton case.}    %

In this section we test our method for the simplest case to treat, that is the one-soliton case, where $N=1$  and the Jost coefficient $b$ is null for 
real eigenvalues $\xi$. The  only  discrete eigenvalue  in the upper half complex plane is
\begin{equation}
\zeta_1=\alpha_1+i \beta_1 \qquad \textrm{with $\beta_1>0$}\,.
\end{equation}
In this case, the explicit expression for the Jost functions are known and quite simple \cite{kaup76}. 
In particular, it is
\begin{equation}
\psi_1(z,t;\zeta_1)=\textrm{e}^{i \zeta_1 t} \frac{d^*}{1+|d|^2}\,,
\end{equation}
\begin{equation}
\psi_2(z,t;\zeta_1)=\textrm{e}^{i \zeta_1 t} \frac{1}{1+|d|^2}\,,
\end{equation}
where
\begin{equation}
d:=-\frac{i C_1}{2 \beta_1}\,\textrm{e}^{2i \zeta_1 t}\,. 
\end{equation}

Autocorrelations for $\Delta$ become (see Equations from \eqref{eq.autocorrelation1 for Delta final} to
\eqref{eq.def of E}):
\begin{equation}
\langle\Delta(\zeta_1,z)\Delta^*(\zeta_1,z') \rangle = 2D\, C(z;\zeta_1,\zeta_1) \delta(z-z')
\end{equation}
and
\begin{equation}
\langle\Delta(\zeta_1,z)\Delta(\zeta_1,z') \rangle = 2D\, E(z;\zeta_1,\zeta_1) \delta(z-z'),
\end{equation}
with
\begin{equation}
C(z;\zeta_1,\zeta_1):=|C_1(z)|^2 \int_{-\infty}^{+\infty}
                                            \left[   |\psi_2(z,t;\zeta_1)|^4
                                                     + |\psi_1(z,t;\zeta_1)|^4
                                            \right] dt 
\end{equation}
and 
\begin{equation}
E(z;\zeta_1,\zeta_1):=C_k^2(z) \int_{-\infty}^{+\infty} 
                 2   \psi^2_2(z,t;\zeta_1)\psi^{2}_1(z,t;\zeta_1)\, dt.
\end{equation}

As a consequence,  Equation\ns\eqref{eq.autocorrelation1 for N discrete} becomes 
\begin{equation*}
 \begin{aligned}
    \langle  N(\zeta_1)  N^*(\zeta_1)\rangle  &= 2D \int_0^\mathcal{L}   C(z;\zeta_1,\zeta_1)\,dz    \\
                    & =2D \int_0^\mathcal{L} |C_1(z)|^2
                                                                       \int_{-\infty}^{+\infty} 
                                                                             \left[  | \psi_2(z,t;\zeta_1)|^4
                                                                                       + |\psi_1(z,t;\zeta_1)|^4
                                                                              \right]
                                                                        dt  \,dz \\
                   &=2D \int_0^\mathcal{L} |C_1(z)|^2 \frac{2 \beta_1}{3 |C_1(z)|^2}            \,dz 
\end{aligned}
\end{equation*}
and therefore
\begin{equation}\label{eq.autocorrelation1 for N discrete 1-soliton}
    \langle  N(\zeta_1)  N^*(\zeta_1)\rangle  
               = \frac{4}{3} D \mathcal{L} \beta_1\,. 
\end{equation}
Similarly, one obtains
\begin{equation*}
 \begin{aligned}
    \langle  N(\zeta_1)  N(\zeta_1)\rangle  &= 2D \int_0^\mathcal{L}   E(z;\zeta_1,\zeta_1)\,dz    \\
                    & =2D \int_0^\mathcal{L} C_1^2(z)
                                                              \int_{-\infty}^{+\infty} 
                                                                    2\psi_1^2(z,t;\zeta_1) \psi_2^2(z,t;\zeta_1)
                                                                dt \,dz \\
                   &=2D \int_0^\mathcal{L} C_1^2(z)\,\left[ - \frac{ \beta_1}{3 C_1^2(z)}\right]    \,dz 
\end{aligned}
\end{equation*}
hence
\begin{equation}\label{eq.autocorrelation2 for N discrete 1-soliton}
    \langle  N(\zeta_1)  N(\zeta_1)\rangle  
               =- \frac{2}{3} D \mathcal{L} \beta_1\,. 
\end{equation}
 The linear growth with the length $\mathcal{L}$ of the fiber is consistent with the analogous linear growth proven in \cite{nature16} in the case of continuous spectrum. 
Furthermore, notice that $\langle  N(\zeta_1)  N(\zeta_1)\rangle$ is real and negative, therefore the real and the imaginary parts of $N(\zeta_1)$ are uncorrelated and the variance of the imaginary part is greater than that of the real one, see also \cite{you14-III}.

These results correspond to the existing literature on the one-soliton case, see for example Chapter 5 of \cite{ian98}. However our method can be in principle applied to arbitrary multiple soliton case, which will be the subject of our future work.

\section{Conclusions.}\label{sec.Conclusions.}   %

The existing networks and infrastructure for telecommunication services are becoming inadequate because the demand  has become increasingly urgent over the years. 
In the late Seventies, the fiber-optic communications were introduced leading to great technological progress that
allowed to exponentially increase the data traffic. 
The Nonlinear Schr\"odinger Equation (NLSE) well models the evolution of the complex envelope of
signals propagating in optical fibers, and its stochastic version \eqref{eq.stochastic NLSE} is necessary to account for the unavoidable presence of noise along the fiber. 
The strong distortion of the optical signal, due to the combined dispersive effects and nonlinear mixing of signal and noise, makes it very difficult to decode the information at the receiver.

The new paradigm of {\em eigenvalue communication}, based on the Nonlinear Fourier Transform, is aimed to exploit the  mathematical integrability of NLSE in order to 
take advantage from nonlinearity, rather than avoiding it.
However, the noise breaks down the perfect integrability of he deterministic NLSE, and this requires an accurate study of the effects of a noisy data on the NFT.
In this paper we have integrated the existing studies on the effect of noise on the continuous part of the nonlinear spectrum with an analogous study on the discrete part.
Assuming the signal-to-noise ratio to be high, we have used a perturbative approach to obtain, at first order, explicit expressions for the propagated statistics 
of the discrete eigenvalues along the fiber. 
Finally, we have tested our results in the mono-solitonic case, that is the case of a single eigenvalue in the upper half complex plane.

\section*{Acknowledgements.}  %
We acknowledge support from Ente Cassa di Risparmio di Firenze, project {\em NOSTRUM - Nonlinear Spectrum Modulation}, ref.\ 2015.0906.


\end{document}